\documentclass{ragtime} 
\usepackage[utf8]{inputenc}
\usepackage{multirow}
\usepackage[subnum]{cases}

\title[Evolutionary tracks of millisecond pulsars]
      {Evolutionary tracks of millisecond pulsars with low-mass companions}

\author[F. Ficek, 
        M. Rutkowski   
        and W. Kluźniak]
       {Filip Ficek\at{1} 
        Mieszko Rutkowski\at[]{1} 
        and Włodek Kluźniak\at[]{2}\\
        \\
        \ins{1}Institute of Physics, Jagiellonian University,\splitins[1] 
         ul. Łojasiewicza 11, 30-348 Krak\'ow, Poland\\         
        \ins{2}Nicolaus Copernicus Astronomical Center, \splitins[1]
        Bartycka 18,
        00-716 Warszawa, Poland\\
}

\begin{document}

\begin{abstract}
 We consider the evolution of millisecond radio pulsars in binary systems
 with a main-sequence or evolved stellar companion.   Evolution of
 non-accreting binary systems with ``eclipsing'' milisecond pulsars
 was described by Klu\'zniak, Czerny \& Ray (1992)
 who predicted that systems like the one containing  the Terzan 5
 PSR 1744-24A will in the future  become accreting low mass X-ray
 binaries (LMXBs), while PSR 1957+20 may evaporate its companion. The
 model presented in the current paper gives similar results for these
 two objects and allows to obtain diverse evolutionary tracks of
 millisecond pulsars with low mass companions (black widows).
 Our results suggest that the properties of many black widow systems
 can be explained by an ablation phase lasting a few hundred million
 years. Some of these sources may regain Roche lobe contact
 in a comparable time, and become LMXBs.
\end{abstract}



\section{Introduction}
Millisecond pulsars are probably intimately connected with
LMXBs, as was realized soon after their discovery:
it was suggested that millisecond pulsars have been spun up
in LMXBs and will end their history in the radio pulsar phase 
\citep{1982CSci...51.1096R,1982Natur.300..728A}.
However, with the discovery of the eclipsing pulsars it was realized
that some millisecond pulsars currently ablating their companions
may re-enter the LMXB phase in a later epoch
\citep{1989Afz....31..567B,1991PAZh...17..433E,Kluzniak}.
Recent discoveries of many ablating binary systems have led to a rekindling
of these ideas, and to the necessity of explaining the evolutionary status
of these black widows and redbacks, as they are called
\citep [e.g.,][]{2014AN....335..313R}.

We are presenting an evolutionary model describing a binary system composed of a
pulsar and its stellar companion. The model includes effects like
gravitational wave emission by the binary, ablation of the companion,
 and pulsar spindown. In general,
part of the ablated matter may accrete onto the neutron
star and another part may leave the system.  The computed evolutionary
tracks begin with the pulsar turn-on at the conclusion of the standard
epoch of accretion in a semi-detached phase.  Throughout most of the
computed evolutionary history, the separation between the pulsar and
the companion star is large enough for the latter to be below its
Roche lobe. Therefore the only mechanism of mass loss considered in
our model is ablation by the pulsar wind.

\section{Model description}
\label{s1}
The period of a binary system including a pulsar of mass $M$
 and its companion of mass $m$ is
\begin{equation} 
	P=\frac{2 \pi J^3 (M+m)}{G^2 M^3 m^3},
\label{eqn:period}
\end{equation}
where $J$ denotes total orbital angular momentum. 
The rate of change of the companion mass $m$ is assumed to be proportional to
the spin-down flux
\begin{equation} 
	\dot{m}\propto \frac{\dot{E}}{4\pi d^2} m^a P^b,
\label{eqn:dmass}
\end{equation}
where $\dot{E}$ is the energy loss of the pulsar primary owing to its spindown,
$d$ is the separation between the primary and the secondary,
 and $a$, $b$ are model dependent exponents. In the simple model assumed
 in \citet{Kluzniak} $a=b=0$. However, in \citet{Brookshaw}
one may find $a=1/6$ and $b=-4/3$. We will adopt the latter values.
 The change of mass of the primary is in principle connected with $\dot{m}$ as
$	\dot{M}=-\beta \dot{m}$.
The coefficient $\beta$ describes how much of the mass lost by the companion
is accreted by the neutron star, and how much is lost from the binary
in a wind, thus $0\le\beta\le1$ with 0 corresponding to no accretion and 1
to no wind.  We will take $\beta=0$.

The change of
angular momentum [first term in Eq.~(\ref{eqn:pp})]
is connected with two processes: emission of the
gravitational waves (GW) and mass loss from system. We take the rate of
angular momentum loss  to gravitational waves to be described by
\citep[e.g.,][]{Shapiro}
\begin{equation} 
	\dot{J}_{GW}=-\frac{256 \pi^3}{5} \frac{G}{c^5} \frac{J^2}{P^3}.
\label{eqn:gw}
\end{equation}
If we assume that specific angular momentum carried away by a wind escaping
 from the system is $j=\alpha M J/[m(m+M)]$, we have
\begin{equation} 
	\dot{J}_{\dot{m}}=\alpha (1-\beta) \frac{MJ}{(M+m)} \frac{\dot{m}}{m}.
\label{eqn:dotj}
\end{equation}
Both  Eqs.~ (\ref{eqn:gw}) and (\ref{eqn:dotj}) conribute to the
 rate of change of the angular momentum:
$  \dot{J}=\dot{J}_{\dot{m}}+\dot{J}_{GW}$. 
By differentiating Eq.~(\ref{eqn:period})
 with respect to time we get the rate of change of the period
\begin{gather}
	\frac{\dot{P}}{P}=
    3\frac{\dot{J}}{J}-\frac{2M+3m}{M+m}
     \frac{\dot{M}}{M}-\frac{3M+2m}{M+m} \frac{\dot{m}}{m},\label{eqn:pp}\\
   \dot m=\gamma\frac{\dot E m^{1/6} P^{-4/3}}{4 \pi  d^2}, \label{eqn:mm} \\
\dot M=-\beta  \dot m,\label{eqn:MM} \\
\frac{\dot J}{J}=\frac{\alpha  (1-\beta ) M}{m+M}\frac{\dot m}{m}-
\frac{256 \pi ^3 G J}{5 c^5 P^3} .\label{eqn:jj}
\end{gather}
Equations (\ref{eqn:pp}), (\ref{eqn:mm}), (\ref{eqn:MM}), (\ref{eqn:jj})
 constitute a system of first-order ordinary differential equations,
 which we proceed to solve with various assumptions and different 
initial conditions.

In the simple case of no accretion onto the primary star, negligible
companion mass, $m<<M$,  and hence negligible gravitational wave
emission, the equations reduce to \citep{Kluzniak} 
\begin{equation} 
	\frac{\dot{P}}{P}=3(\alpha-1)\frac{\dot{m}}{m},
\label{eqn:dotpkluzniak}
\end{equation}
and can be easily integrated, yielding
\begin{equation} 
	P(m) \propto m^{3(\alpha-1)}.
\label{eqn:logkluzniak}
\end{equation}
With suitable initial conditions the evolutionary paths on the
 $P$ vs. $m$ plot described by this equation can be made to pass
 through the current positions
of some of the known pulsars, e.g. PSR 1957-20 (see the Appendix, Fig.~2).

The source of the energy driving the ablation process is pulsar spindown.
 From the magnetic dipole formula \citep[e.g.,][]{Shapiro} we have
\begin{equation} 
	\dot{E}=-\frac{B^2 R^6 \Omega^4 \sin^2 \theta}{6c^3},
	\label{eqn:slowdown}
\end{equation}
where $B$ denotes the surface magnetic field near the pole, $\Omega$ is
the pulsar spin rate (the pulsar period being ${P_0}={2\pi}/\Omega$), $R$ is
the pulsar radius and $\theta$ denotes the angle between the magnetic
and the rotation axes (for simplicity we take $\sin^2\theta=1$
 and $R=10\,$km). On the other hand we have 
$\dot{E}=I\Omega \dot{\Omega}$, where
$I$ is the moment of inertia of the pulsar. These two equations provide
\begin{equation} 
	\Omega(t)=\frac{\Omega_{0}}{\sqrt{2t/\tau+ 1}},
	\label{eqn:omega}
\end{equation}
where $\Omega_{0}=\Omega(0)$ is the initial angular velocity of the
pulsar and $\tau=-\Omega(0)/\dot\Omega(0)$
 is the characteristic age of the pulsar (at time t=0).
Eqs.~(\ref{eqn:slowdown}),
(\ref{eqn:omega}), are used in Eq.~(\ref{eqn:mm}) to find
$\dot{m}$ as a function of time.

\begin{figure}[!h]
\begin{center}
\includegraphics[width=\linewidth,height=1.0\linewidth]{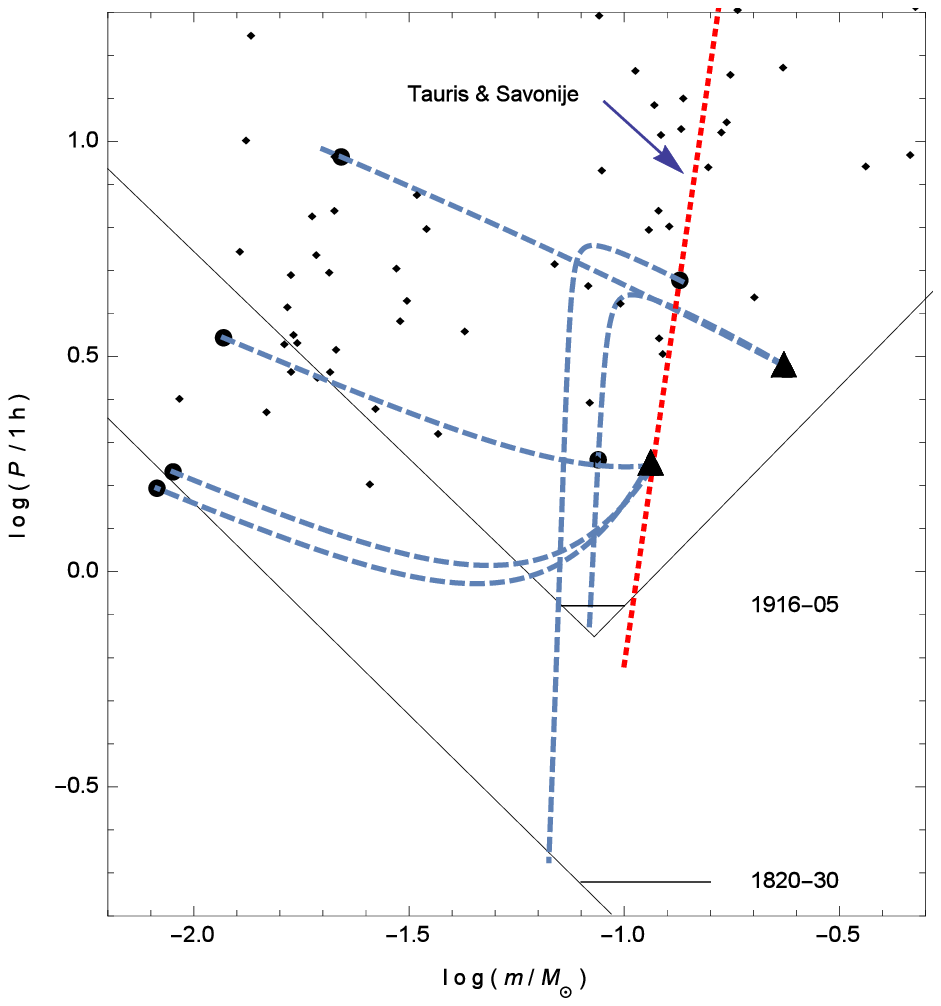}
\caption{\label{fig:wykres1}
Large dots correspond to the present parameters of the observed
pulsar systems for which the evolutionary tracks have been computed
(dashed blue lines). Small dots are other objects taken from the
\cite{Atnf}.
Solid lines correspond to Roche lobe contact for a cold
companion.  The short-dashed red line,
 taken from \citet{Tauris}, corresponds to LMXB evolution
of a system with an evolved companion.
Also shown (thin horizontal line segments) are the positions
of two short-period LMXBs. The filled triangles mark plausible initial
points of the evolutionary tracks.}
\end{center}
\end{figure}

When the secondary star is sufficiently close to the pulsar that it fills
the Roche lobe, accretion through the inner Lagrangian point
starts. This situation is not described by our model, although our
tracks may bring the system to this point.  When the radius of the
companion is equal to Roche lobe radius, the relation between orbital
period $P$ and companion mass $m$ is
\begin{equation} 
	P=2\pi \sqrt{\frac{A^3}{B^3 G}} m^{(3n-1)/2},
	\label{eqn:paczynski}
\end{equation}
where $B\approx0.462$. The values of $n$ and $A$ correspond to the
radius of the companion thorugh $r=A m^n$.
 For degenerate stars like white dwarfs,
$n=-1/3$, and for a hydrogen white dwarf $A=2.82\cdot
10^4M^{1/3}_\odot\,$km  \citep{Shapiro}, while from \cite{Hamada} one
obtains $A=8.80\cdot 10^3 M^{1/3}_\odot\,$km for a helium white dwarf.
One may also obtain this coefficient for a carbon white dwarf, which
is $A=8.72\cdot 10^3 M^{1/3}_\odot\,$km  \citep{Hamada}, it is almost
indistinguishable from the helium one. Lines corresponding to
Eq.~(\ref{eqn:paczynski})
indicate where
 the evolutionary track may terminate in a Roche-lobe
overflowing LMXB, depending on the companion type
(Fig. \ref{fig:wykres1}).

\section{Results of numerical calculations}
\label{conclus}

Using Mathematica, we solved numerically the system of four
differential equations, Eqs.~(\ref{eqn:pp}), (\ref{eqn:mm}),
(\ref{eqn:MM}), (\ref{eqn:jj}) discussed above.
We consider a model with no accretion ($\beta=0$), we assume
$\gamma=2.5\cdot10^4\,{\rm s}^{10/3}\,{\rm g}^{-1/6}$
 \citep[cf.,][]{2013ApJ...775...27C} and,
 following \citet{Kluzniak}, we take
$\alpha=0.86$.    For the initial point on
the ($m$, $P$) plane we use one of two points on the track of
\citet{Tauris}, which describes the evolution of a LMXB with an
evolved companion. For the PSR 1957+20 and B1744-74A (Terzan 5) tracks
we use the starting point of \cite{Kluzniak}, corresponding to the point
at which magnetic braking is supposed to lose importance in the evolution
of binaries with a main sequence companion.  Current system parameters
are taken from \citet{Atnf,2005AJ....129.1993M}, and they can be found
 in Table~1, together with other data,
 for the six tracks which are presented in Fig.~1.


Derived times of evolution are $t_{\rm ev}\simeq7\cdot10^8\,$y for PSR
1957+20 and $t_{\rm ev}\simeq5.5\cdot10^9\,$y for Terzan 5. For PSR
1957+20, evolution is steady, whereas for Terzan 5 one can distinguish
three stages of evolution. The first stage, when the evolution curve is
nearly a straight line, lasts about $4.5\cdot10^8\,$y.  The second one,
when the evolution path ``turns downwards" on the $P$ vs $m$ plot, lasts
$2.4\cdot10^9\,$y. The last stage, when gravitational radiation is
dominant, lasts $2.6 \cdot 10^9\,$y.  Objects with convex evolution
curves evolve comparably fast: e.g. for J1807-2459A the evolution time is
 $t_{\rm  ev}\simeq5.7\cdot10^8\,$y.
The values of $t_{\rm  ev}$ in parentheses in Table~1 (for the Terzan 5
pulsar and  J1023+0038) correspond to the time it will take for the system
to regain the line of Roche-lobe contact starting from the present position.

\begin{table}
\label{tab:parameters}
\begin{center}
\begin{tabular}{ccccc}
\hline
\multirow{2}*{Quantity}&\multicolumn{2}{|c|}{\textbf{PSR 1957+20}}&\multicolumn{2}{|c|}{\textbf{B1744-74A/Terzan 5}}\\
\cline{2-5}
&\small{Initial}&\small{Present}&\small{Initial}&\small{Present}\\
\hline
$P$[hr]&2.9&9.2&3.0&1.82\\
\hline
$m[M_\odot]$&0.235&0.022&0.235&0.087\\
\hline
$M[M_\odot]$&1.7&1.7&1.4&1.4\\
\hline
$\dot m$[g/s]&$-2.0\cdot10^{17}$&$-7.6\cdot10^{14}$&$-1.1\cdot10^{17}$&$-3.1\cdot10^{14}$\\
\hline
$P_0$[ms]&0.92&1.60&1.95&11.56\\
\hline
$B$[G]&\multicolumn{2}{|c|}{$4.0\cdot 10^{8}$}&\multicolumn{2}{|c|}{$1.3\cdot10^{9.0}$}\\
\hline
$\mu$[G$\cdot$ cm$^3$]&\multicolumn{2}{|c|}{$4.0\cdot10^{26}$}&\multicolumn{2}{|c|}{1.3$10^{27}$}\\
\hline
$t_{\rm ev}$[y]&\multicolumn{2}{|c|}{6.72$\cdot$10$^8$}&\multicolumn{2}{|c|}{5.52$\cdot$10$^9$ (5.90$\cdot$10$^9$)}\\
\hline

\multirow{2}*{Quantity}&\multicolumn{2}{|c|}{\textbf{J1807-2459A}}&\multicolumn{2}{|c|}{\textbf{J2241-5236}}\\
\cline{2-5}
&\small{Initial}&\small{Present}&\small{Initial}&\small{Present}\\
\hline
$P$[hr]&1.75&1.71&1.75&3.50\\
\hline
$m[M_\odot]$&0.115&0.009&0.115&0.012\\
\hline
$M[M_\odot]$&1.4&1.4&1.4&1.4\\
\hline
$\dot m$[g/s]&$-4.5\cdot10^{15}$&$-2.6\cdot10^{15}$&$-1.4\cdot10^{16}$&$-1.1\cdot10^{15}$\\
\hline
$P_0$[ms]&2.91&3.06&1.98&2.19\\
\hline
$B$[G]&\multicolumn{2}{|c|}{$2.9\cdot 10^{8}$}&\multicolumn{2}{|c|}{$2.4\cdot10^{8}$}\\
\hline
$\mu$[G$\cdot$cm$^3$]&\multicolumn{2}{|c|}{$2.9\cdot10^{26}$}&\multicolumn{2}{|c|}{$2.4 \cdot 10^{26}$}\\
\hline
$t_{\rm ev}$[y]&\multicolumn{2}{|c|}{5.60$\cdot$10$^8$}&\multicolumn{2}{|c|}{7.88$\cdot$10$^8$}\\
\hline

\multirow{2}*{Quantity}&\multicolumn{2}{|c|}{\textbf{J1311-3430}}&\multicolumn{2}{|c|}{\textbf{J1023+0038}}\\
\cline{2-5}
&\small{Initial}&\small{Present}&\small{Present}&\small{Predicted}\\
\hline
$P$[hr]&1.75&1.56&4.73&0.31\\
\hline
$m[M_\odot]$&0.115&0.008&0.136&0.061\\
\hline
$M[M_\odot]$&1.4&1.4&1.4&1.4\\
\hline
$\dot m$[g/s]&$-3.6\cdot10^{15}$&$-3.0\cdot10^{15}$&$-2.2\cdot10^{16}$&LMXB\\
\hline
$P_0$[ms]&2.48&3.56&1.67&11.09\\
\hline
$B$[G]&\multicolumn{2}{|c|}{$2.0\cdot 10^{8}$}&\multicolumn{2}{|c|}{$7.9\cdot10^8$}\\
\hline
$\mu[G\cdot cm^3]$&\multicolumn{2}{|c|}{$2.0\cdot10^{26}$}&\multicolumn{2}{|c|}{$7.9\cdot10^{26}$}\\\hline
$t_{\rm ev}$[y]&\multicolumn{2}{|c|}{5.44$\cdot$10$^8$}&\multicolumn{2}{|c|}{(1.8$\cdot$10$^{10}$)}\\
\hline

\end{tabular}
\caption{System parameters}
\end{center}
\end{table}
\centerline {}

\section{Evolutionary tracks}\label{s2}

Evolution of the system depends on the ratio between angular momentum
losses caused by ablation and gravitational wave emission.  There
seem to be three types of tracks. 

In the case where gravitational waves emission can be neglected (like
in the PSR 1957-20 system) the track is well described by the formula of
Eq.~(\ref{eqn:logkluzniak}). The system very nearly follows a straight
line on a $\log P$ versus $\log m$ plot. The slope of this line
depends only on the parameter $\alpha$. The track may be deflected 
a little bit due to vestigial gravitational wave emission.

Another possible track passes through the Terzan 5 pulsar
B1744-24A. In the initial phase of system evolution the track is
similar to the one described in the previous paragraph. The difference
is that at a certain moment, owing to pulsar spindown,
gravitational wave emission starts to dominate over ablation. If, from
that point on, mass loss were neglected (i.e., the  evolution were
driven by GW emission alone), the track would be a vertical line on
the  $\log P$--$\log m$ plot. In fact, a residual effect of ablation
is still felt, and the track deviates slightly in the direction of
lower companion mass (to the left in the figures).

Neglecting mass loss from and mass transfer in
the system ($\gamma=0$ in Eq.~[\ref{eqn:mm}])
 one easily obtains the time
elapsed in the evolution from binary period $P_{i}$ to period $P$:
\begin{equation} 
	T=\frac{5 c^5}{2048 \pi^3 G J_i} (P^{8/3} P_{i}^{1/3}-P_{i}^3),
\label{eqn:spindown}
\end{equation}
where $J_i$ is the initial angular momentum (corresponding to $P_{i}$).
 Time scales of evolution obtained from this equation are similar
 to the numerical values for the nearly vertical tracks in Fig.~1.

Tracks similar to those described above were already obtained by
\citet{Kluzniak}. They cover situations in the limit where one of the
effects, ablation or GW emission, dominates over the other along each
major segment of the trajectory
 (although, as remarked above in Section~\ref{conclus}, PSR B1744-74A 
 spends most of its evolutionary time in transition between
two such states).  It seems that systems with an evolved very low mass
companion ($m<0.04M_\odot$) cannot evolve this way. For instance,
obtaining a ``Terzan-like'' evolution track for these systems leads to
evolution time amounting to several dozens of billion years.  A third
type of evolutionary track seems to be required. 

 We have found evolutionary tracks connecting the currently
observed binary parameters of the pulsars J2241-5236, J1807-2459A
and J1311-3430
with a plausible inital point and having reasonable
time scales of evolution. These evolutionary tracks are characterized by
angular momentum loss to  both GW emission and ablation effects, and
have a convex shape on a log $P$ versus log $m$ plot
(Fig.~\ref{fig:wykres1}).  Eventually, the separation of the system
components becomes large enough that GW emission loses importance, and
  the track becomes parallel to that of PSR 1957+20.

\section{Discussion}\label{dsc}
We have considered the evolution of millisecond radio pulsars with
binary low-mass companions assuming simple formulae for the
ablation rate of the companion by the pulsar wind.
For the starting point of each evolutionary track that we considered
we have taken a plausible moment of pulsar turn-on in an erstwhile LMXB,
either along the standard evolutionary curve familiar from  discussion
of cataclysmic variables and the period gap,
i.e., a binary with a main-sequence companion
\citep{1983ApJ...268..825P}, or along an evolutionary
track with an evolved companion
 \citet{Tauris}.
Pulsar turn-on (or turn-off) in (potentially) accreting low mass
binaries was discussed in \citet{1988Natur.334..225K}.

We have reproduced the results of \citet{Kluzniak}
who performed a similar study for the only two known
eclipsing pulsars at the time (PSR 1957+20 and B1744-74A in Ter 5),
and found that there are periods of their evolutionary history in the ablation
phase when either one or the other of two major angular momentum loss
mechanisms  dominates (mass loss from the system or GW emission).
We note that evolutionary tracks that we now find based on
the \citet{Brookshaw} evaporation formula, Eq.~(\ref{eqn:dmass}),
imply shorter initial pulsar periods than previously obtained,
this can be seen from a comparison of the entries in Table 1 with 
the description of tracks (b) and (e) in the Appendix, Fig.~2.

We find that we are able to reproduce the current positions
of typical millisecond radio pulsars with a low mass binary
companion, typically this involves an ablation phase lasting several
hundred years. However, 
we find that for the majority of the black widow pulsars
known today the relative importance of the two considered angular momentum
loss mechanisms is comparable in their evolutionary history,
i.e., unlike in the case of
PSR 1957+20 and B1744-74A, neither GW emission nor mass loss
domiantes the other over major portions of the evolutionary track
in the period-mass diagram (Fig.~1).

We confirm the conclusion of
\citet{Kluzniak}, who predicted that some ms pulsars may become
accreting LMXBs at the end of their evolution.
Two of the tracks presented in this paper end very close
to the line of Roche-lobe contact, in the current position of
PSR J1807-2459A and PSR J1311-3430. These two pulsars
seem to be close to the end of a $5\times10^8\,$y ablation phase.

We note that detailed binary evolutionary calculations, which included
an ablation phase similar to the model considered here
were presented recently in \citet{2013ApJ...775...27C}.

\ack
We thank Dr. Thomas Tauris, as well as the anonymous referee,
 for many detailed comments on the manuscript.
This work was supported in part by NCN grant 2013/08/A/ST9/00795.



\bibliography{pftrackRev}
\newpage
\section{Appendix}
For ease of reference, we reproduce Figure 1 and its caption
from the pre-arXiv contribution of Klu\'zniak, Czerny and Ray (1992).

\begin{figure}[!h]
\begin{center}
\includegraphics[width=\linewidth,height=0.7\linewidth]{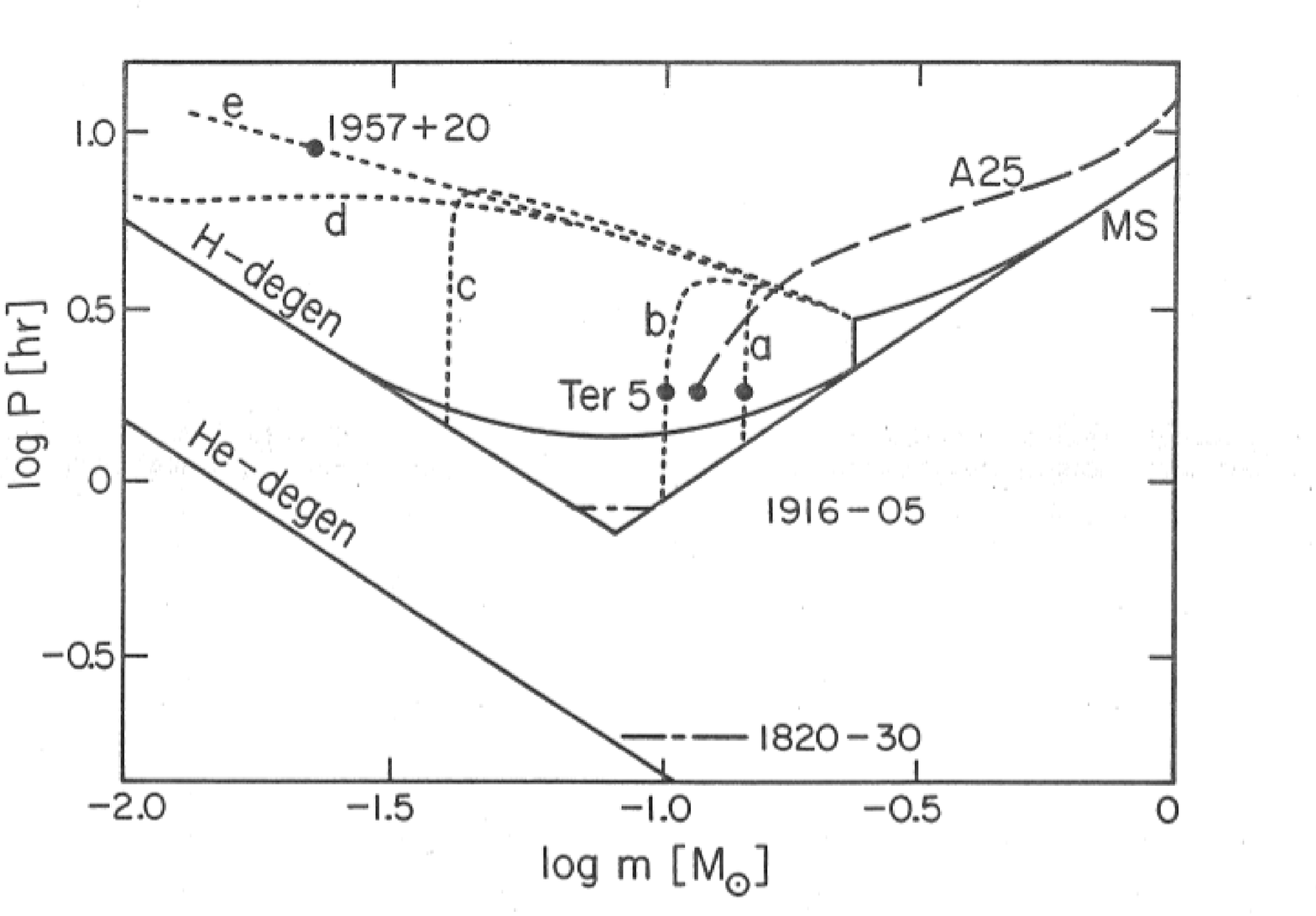}
\caption{
 `Figure 1. Possible evolutionary tracks of systems with ``evaporative''
  mass loss.  The decimal logarithm of the orbital period in hours is
  plotted versus the decimal logarithm of the mass of the companion in
  units of Solar mass. Likely location of the eclipsing pulsars (filed
  circles) as well as possible positions of the X-ray binaries
  4U$\,$1916-05 and 4U$\,$1820-30 are also indicated (dash-dot-dash lines).
  The thick straight line segments correspond to systems with a
  main-sequence or a cold degenerate dwarf companion in Roche-lobe
  contact. According to the standard theory of their evolution,
  cataclysmic variables follow the thin curve (in the direction of
  decreasing companion mass, $m$). When this theory is applied to
  canonical LMXBs, the dotted tracks ensue, see Section 5 for
  details. The lines (a) through (e) differ only in the properties of
  the pulsar ablating its companion: in the strength of the magnetic
  dipole moment and in the initial value, $P_0$, of the rotational period
  of the neutron star. The values of $P_0$ and $\log(B/$Gauss), where
  $B\equiv\mu\times 10^{-18}\,$cm$^{-3}$, are respectively (a) 5.0 ms,
  9.5; (b) 3.4 ms, 8.9; (c) 2.0 ms, 9.0; (d) 2.0 ms, 8.6; (e) 1.25 ms,
  8.1. We assumed that 10\% of the energy flux impinging on the
  companion is converted into kinetic energy of the evaporative
    plume, and we took $\beta=0.86$.' }
\label{fig:czerny}
\end{center}
\end{figure}
\par\noindent {N.B. The parameter ``$\beta$'' in the quoted caption
corresponds to our $\alpha$.\hfill}

\end{document}